# Performance Comparison on Parallel CPU and GPU Algorithms for Unified Gas-Kinetic Scheme

Jizhou Liu[(1)], Fang Q. Hu[(2)] and Xiaodong Li[(1)]

(1) *School of Energy and Power Engineering, Beihang University, Beijing 100191, China*
(2) *Department of Mathematics and Statistics, Old Dominion University, Norfolk, VA 23529, United States*

**Abstract**: Parallel algorithms on CPU and GPU are implemented for the Unified Gas-Kinetic Scheme and their performances are investigated and compared by a two dimensional channel flow case. The parallel CPU algorithm has a one dimensional block partition that parallelizes only the spatial space. Due to the intrinsic feature of the UGKS, a compromised two-level parallelization is adopted for GPU algorithm. A series of meshes with different sizes are tested to reveal the performance evolution of the algorithms with respect to problem size. Then special attentions are paid to UGKS applications where the molecular velocity space range is large. The comparison confirms that GPU has relative elevated accelerations with the latest device having a speedup of 118.38x. Parallel CPU algorithm, on the contrary, might provide better performances when the grid point number in velocity space is large.

## 1. Introduction

The Unified Gas-Kinetic Scheme (UGKS) is a direct flow modeling of all regimes [1,2]. Based on an integral solution of the BGK model, the scheme constructs a cell interface flux compromising particle collision and free transport mechanism. Therefore, gas dynamics from continuum limit to rarefied regime could be reconstructed automatically depending on the Knudsen number of the flow. This special property makes UGKS a powerful tool in simulating non-equilibrium microscale flows [3] and supersonic high speed flows [4].

The UGKS is a mesoscopic method that involves molecular velocity space. In detail, flow variables and corresponding distribution function are stored and updated in the formulation, which makes it computationally intensive as compared to conventional N-S flow solver. Recently, by the analogous dynamics of rarefied gas with particles, UGKS has been extended to solving gas-particle multiphase coupling problems [5]. Due to the fact that a typical particle's mass is orders heavier than that of a gas molecular, its characteristic velocity $c = \sqrt{2kT/m}$ is much smaller than the sound speed of gas medium. Therefore, a small fluctuation in gas phase would result in wide range of variation in particle velocity space. To maintain accurate integration in particle velocity space, large interval and more grid points are needed thus further slowing down this method.

To improve the computational speed, MPI parallelization on CPU chips has been investigated by Li et al. [6] and Ragta et al. [7] for three dimensional UGKS. In Li's study, both the physical and velocity space are parallelized. Two dimensional Cartesian topology is used to arrange the physical and velocity blocks. Tests on small-scale and large-scale grids show quasi-linear speedups. In Ragta's work, parallelized UGKS on Cartesian grid is constructed and tested on very large problems. Canonical turbulence at low Knudsen numbers are correctly simulated with parallelized UGKS code.

Beside MPI algorithm on CPU, parallel computing with Graphics Processing Unit (GPU) has become an evolutionary solution for large scale calculation in image processing and biomolecule analysis. It has also been applied to the field of Computational Fluid Dynamics. Depending on methods used and size of problem, speedups ranging from 10x to 100x could be reached by implementing GPU programming [8,9]. Karantasis et al. [10] implement a highly accelerated finite difference WENO scheme on GPU card. A maximum 50x speedup is obtained which shows the promising potential of GPU in solving turbulent flow at high Reynolds number. Lou et al. [11] establish a GPU accelerated p-multigrid discontinuous Galerkin (DG) method based on OpenACC for 3D unstructured grid. The application of their method on real engineering problems shows a speedup ranging from 5x to 10x. Besides, GPU acceleration has also been implemented to kinetic methods that treat Boltzmann equation. Boroni et al. [12] builds a fully parallel GPU implementation of LBM in combination with Immersed Boundary Conditions. A 50x to 100x speedups are obtained for two dimensional benchmark test cases.

The Unified Gas-Kinetic Scheme is also a kinetic-based method that needs acceleration techniques before being applied to treat real engineering problems. Till now, efforts have been devoted to its MPI parallelization on CPU clusters. However, no work has been done in trying to implement UGKS on GPU accelerator. Therefore, this paper intends to build a primitive parallel algorithm on GPU for UGKS and compare its performance with parallelized CPU algorithm. To do this, parallel MPI code and CUDA Fortran code are developed for two dimensional UGKS. Their performances are investigated on a benchmark two dimensional channel flow problem with two versions of CPUs and GPUs and 4 kinds of mesh sizes. In view of solving gas-particle multiphase coupling problem with UGKS, computing performances on function of particle velocity space discretization are also tested.

## 2. Parallel implementations of UGKS on CPU and GPU

The Unified Gas-Kinetic Scheme employed in this paper adopts a two dimensional formulation with no external forces. The starting governing equation is the Boltzmann equation with BGK operator.

$$\frac{\partial f}{\partial t}+u\frac{\partial f}{\partial x}+v\frac{\partial f}{\partial y}=\frac{g-f}{\tau} \quad , \quad g(t,x,y,U,V)=\rho\frac{m}{2\pi kT}e^{-\frac{m}{2kT}((u-U)^2+(v-V)^2)} \quad (1)$$

where $f(t,x,y,u,v)$ is the density distribution function in time, physical space and molecular velocity space. $g$ represents the Maxwellian distribution function in equilibrium state.

The UGKS is based on a finite volume method. The physical domain is partitioned into $I \times J$ cells with cell number $(x_i, y_j)$. For each cell, the molecular velocity space is further discretized into $K \times M$ grid points. By method of characteristics, the density distribution function at cell interface $f_{i+1/2,j,k,m}$ could be constructed with an integral solution of Eq. (1) [1]

$$f_{i+1/2,j,k,m}=\frac{1}{\tau}\int_{t^n}^{t}g(t',x',y',u_k,v_m)e^{-(t-t')/\tau}dt'+e^{-(t-t^n)/\tau}f_{0,k,m}^n(x_{i+1/2}-u_k(t-t^n),y_j-u_m(t-t^n),u_k,v_m) \quad (2)$$

the corresponding macroscopic flux $F_{i+1/2,j}$ could be obtained by integrating cell interface distribution in molecular velocity space. Finally the time evolutions of the density distribution function and macroscopic flow variables are

$$f_{i,j,k,m}^{n+1}=f_{i,j,k,m}^n+\frac{1}{dx}\int_{t^n}^{t^{n+1}}u_k(f_{i-1/2,j,k,m}-f_{i+1/2,j,k,m})dt+\frac{1}{dy}\int_{t^n}^{t^{n+1}}u_v(f_{i,j-1/2,k,m}-f_{i,j+1/2,k,m})dt$$

$$+\frac{dt}{2}\left(\frac{g^{n+1}-f_{i,j,k,m}^{n+1}}{\tau^{n+1}}+\frac{g^n-f_{i,j,k,m}^n}{\tau^n}\right) \quad (3)$$

$$W_{i,j}^{n+1} = W_{i,j}^{n} + \frac{1}{dx}(F_{i-1/2,j} - F_{i+1/2,j}) + \frac{1}{dy}(F_{i,j-1/2} - F_{i,j+1/2}) \tag{4}$$

The detailed construction of the cell interface flux could be found in [1,2] and will not be further outlined in this paper.

### A. Parallel block partition for MPI algorithm

Since a simple two dimensional rectangular domain with relatively small number of cells is targeted in this paper, a one dimensional way of domain splitting is adopted for the MPI code to guarantee minimum data transfer. The physical region is sliced into N blocks in y-direction as is shown in Fig.1. The UGKS uses a second order interpolation in physical space, therefore only the data of the leftmost and rightmost column in each block (blue region) need to be transferred between adjacent blocks.

In this block layout, every CPU thread is only responsible for the cells within one MPI-block, which makes it faster than the single-thread sequential calculation. However, if too many blocks are utilized, the time consumed by data transfer would take too big part of the total calculation time. One thing to be noticed is that the molecular velocity space $(u_k, v_m)$ is not parallelized under this configuration.

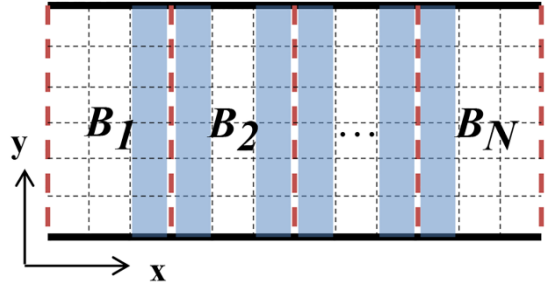

Figure 1 Block partition of MPI algorithm

### B. Block and thread layout for GPU

A typical general purpose GPU card has thousands of threads (or cores) that run in parallel. These threads could be grouped, in a higher hierarchy, into thread blocks that share a small amount but very fast memory spaces (shared memory). The structure of the block and thread is prefixed to be three dimensional. However, its practical usage depends much on the size and structure of the investigated problem.

By analyzing the characteristics of UGKS formulation, a two-level parallelization strategy is adopted herein with the following considerations:

(1) Keeping all available GPU cores busy: use full parallel computing when possible; sequential computing is used for molecular velocity space level integration that demands parallel reduction to avoid idling threads;

(2) Using on-chip shared memory for variables that are in common for the whole thread block;

(3) Prioritizing more thread blocks with smaller size to smaller thread blocks with more threads inside it.

The first parallel strategy is a full parallelization of spatial space and molecular velocity space as is illustrated in Fig. 2(a). Each control volume in physical space is mapped onto one GPU block $B_{i,j}$. Within each thread block, $K \times M$ threads are allocated for each discrete velocity space $(u_k, v_m)$. Considering that smaller thread block might work faster, for each control volume, the velocity grid is further divided into $n$ blocks in $v$ direction. Finally GPU block partition is three dimensional $B_{i,j,n}$ with $K \times (M/n)$ threads inside it. This block-thread allocation pattern explores the full power of GPU device and works efficiently when the discrete velocity space is not coupled from each other. For example, in calculating the density distribution function $f_{i,j,k,m}^{n+1}$ in Eq. (3), the value at each velocity space grid point could be added and multiplied independently. This configuration is used for computing the slopes, interpolations, and updates of density distribution function in GPU code for UGKS.

However, in the process of constructing the macroscopic and microscopic fluxes in UGKS, the aforementioned block-thread partition could not obtain optimal speedups, which is due to the need of parallel reduction and significant fraction of sequential calculations. The major steps in flux evaluation of UGKS are detailed as follow.

| For every cell interface $(x_{i+1/2}, y_j)$ | | |
|---|---|---|
| S1. Evaluate interface flow variable by moment of distribution function: $W_{i+1/2,j}^0 = \sum f_{i+1/2,j}(u_k, v_m)$ | reduction needed x8 | parallel |
| S2. Calculate velocity moments: $Mu$, $Mu_L$, $Mu_R$ (normal), $Mv$ (tangential), $Mxi$ (internal) | | sequential |
| S3. Calculate time related variables: $Mt$, $\tau$ | | sequential |
| S4. Calculate microscopic slope coefficients: $a_L, a_R$ (normal), $b$ (tangential), $A_t$ (time derivative) | | sequential |
| S5. Evaluate hydrodynamic part of the macroscopic flux: $F_{i+1/2,j}$ | | sequential |
| S6. Calculate Maxwellian distribution at interface: $G(u_k, v_m)$ | | parallel |
| S7. Evaluate the free transport part of the macroscopic flux: $F_{i+1/2,j}$ | reduction needed x4 | parallel |
| S8. Evaluate the flux for distribution function: $\tilde{f}_{i+1/2,j}(u_k, v_m)$ | | parallel |

The parallel strategy in Fig.2 (a) is only efficient for step S6 and S8 where each discretized velocity is fully independent. For Step S1 and S7, the moment calculation of the distribution function involves integration in velocity space. In this process, shared memory based parallel reduction is needed which deteriorates the efficiency. The reason for the deterioration is that while doing parallel reduction in shared memory, idling thread issue is unavoidable. These threads in idle state could be otherwise allocated to other jobs instead of waiting synchronization calls. Another fact is that during the flux evaluation in UGKS, there are a total of 12 times parallel reduction needed which makes the idling thread issue more significant. Last but not least, there are large parts of the code that would be sequential in the configuration of Fig. 2 (a), which also involves thread idling problem if full parallelization is applied.

To keep all computing resources busy, a compromise is made in the flux evaluation process for UGKS. For step S6 and S8, full spatial space and velocity space parallelization in Fig. 2(a) is adopted. For the other steps, a second spatial space level parallelization is used as illustrated in Fig. 2 (b). In detail, block partition is one dimensional with each thread $T_{k,m}$ mapped onto one control volume. Here, the flux evaluations at different cell interfaces are done in parallel while the whole calculation on individual cell interface is done in sequential order. This concerns a competing relationship between doing moment integrations in sequential order and in parallel with idling threads. It has been tested that using sequential computing for step S1 and S7 is faster than using parallel computing with reduction.

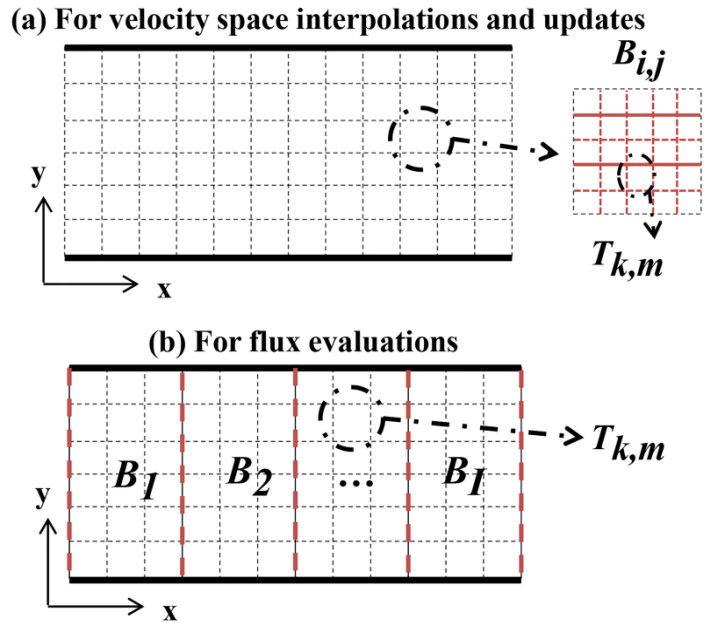

Figure 2 Block and thread partition of GPU algorithm

## 3. Test case and result analysis

### A. Test case description

The studied case herein is a 2 dimensional micro channel flow with pressure gradient. The channel length/width ratio is 2. The Knudsen number calculated with the channel width as the characteristic length is Kn = 0.1. The flow in the channel is driven by given pressure at inlet and outlet. For this case, with a reference pressure of $p_0$, the inlet and outlet pressures are $p_+ = 1.2 p_0$ and $p_+ = 0.8 p_0$ respectively.

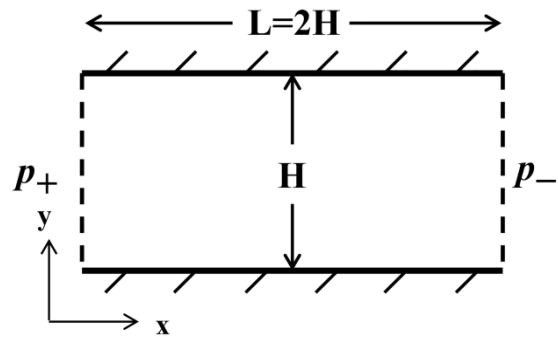

Figure 3 Two dimensional channel with pressure gradient

The computational region is discretized into Cartesian grid of $I \times J$ control volumes. 4 different mesh sizes are tested ranging from 800 cells to 12800 cells. These numbers seems small as compared to other CFD code tests with thousands or millions of grid points. The fact is that UGKS is a mesoscopic method. In its molecular velocity space, 3 kinds of discretization are also adopted with increasing density. Finally, the total degree of freedom ranges from 676,000 to 60,992,000, which is not a small number.

Table 1 Spatial mesh partition and velocity space discretization

| Spatial mesh | I | J | Velocity space discretization | |
|---|---|---|---|---|
| M1(40×20) | 40 | 20 | K | M |
| M2(80×40) | 80 | 40 | 29 | 29 |
| M3(120×60) | 120 | 60 | 49 | 49 |
| M4(160×80) | 160 | 80 | 69 | 69 |

The CPU computing resources come from Fluid and Acoustic Engineering Laboratory, Beihang University, China and the GPUs are provided by the Research Computing Service of Old Dominion University, U.S.A. Two versions of CPU processors and GPU processors are utilized. The following table summarizes the main parameters of the CPUs and GPUs.

Table 2 Basic parameters of CPUs and GPUs

| Processor Type | CPU | | GPU | |
|---|---|---|---|---|
| Version | Intel-E5-2650 v1 | Intel-E5-2695 v3 | Nvidia Tesla K-40 | Nvidia Tesla V-100 |
| Core number | 8 cores, 16 threads | 18 cores, 36 threads | 2880 | 5120[*] |
| Clock rate | 2.0GHz | 2.3GHz | 0.875GHz | 1.5GHz |
| Memory | 32GB | 32GB | 12GB | 16GB[*] |
| Compute ability | --- | --- | 3.5 | 7.0 |
| Year of production | 2011 | 2016 | 2013 | 2017 |

* The Nvidia V-100 device usage is collective. Jobs from other users may be running simultaneously

### B. Results and analysis

The test case is run on CPUs and GPUs with sequential and parallel codes. The single-thread sequential results serve as references in calculating parallel speedups. In MPI calculations, the maximum block number is set to $J/2$ for E5-2650 version CPU and $J$ for E5-2695 version CPU. The channel flow simulation is run for 10,000 steps for all cases. Identical results are obtained on both CPUs and GPUs for the same mesh size.

*Speedups with mesh sizes*

First, we fix the discretization in molecular velocity space to $K\times M = 29\times 29$ and run the codes on meshes of different sizes. Fig. 4 plots the speedups for the CPU-MPI code on the two versions of CPU. Generally, the speedups follow the Amdahl's law, which means they increase with mesh size and number of threads used. At a certain thread number, the speedup lines reach a plateau meaning that further addition of CPU threads will not eventually increase the speed of calculation for UGKS. The maximum speedups for Mesh 4 are 11.64x on E5-2650 and 15.23x on E5-2695 device. One thing to be noticed is that the speedup plots fall quickly into sublinear region, which might be due to the relative small size of the problem.

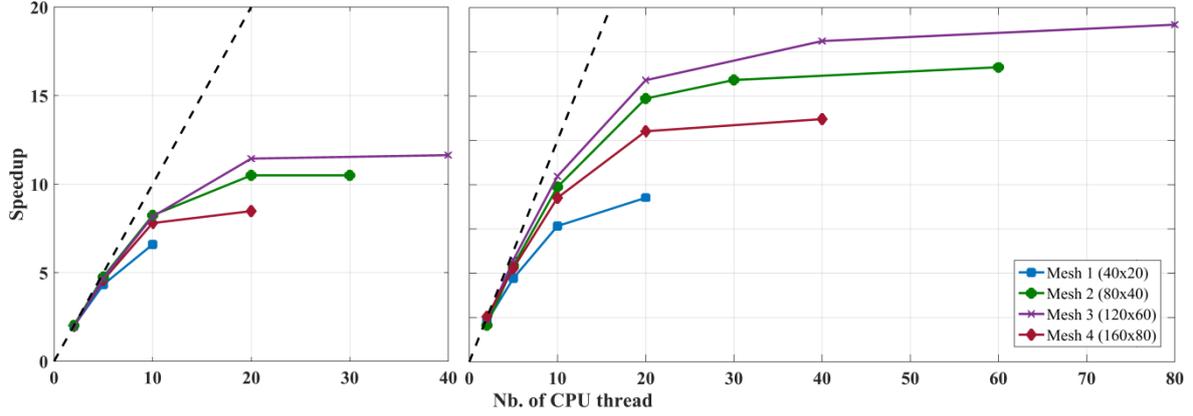

Figure 4 Speedups of MPI code on two versions of CPU

Table 3 presents the computation times of single and parallel algorithms on both CPUs and GPUs. To compare more directly their performances, Fig. 5 plots the speedups for CPUs and GPUs on function of mesh size. The reference time here is set to be the computing time of the single thread E5-2650 CPU and the reference mesh size is the size of Mesh 1 in Table 1.

Table 3 Computation times for parallel CPU and GPU algorithms

| Spatial mesh | CPU: E5-2650 | | CPU: E5-2695 | | GPU | |
|---|---|---|---|---|---|---|
| $I \times J$ | 1 thread | $J/2$ thread | 1 thread | $J$ thread | K-40 | V-100 |
| M1(40×20) | 1749.13 s | 265.12 s | 1004.63 s | 135.55 s | 141.77 s | 46.68 s |
| M2(80×40) | 8507.09 s | 1003.10 s | 4351.73 s | 396.82 s | 451.21 s | 93.29 s |
| M3(120×60) | 19603.08 s | 1866.27 s | 9849.67 s | 739.98 s | 1216.52 s | 169.99 s |
| M4(160×80) | 34076.52 s | 2926.90 s | 18040.99 s | 1184.41 s | 2486.85 s | 287.83 s |

A common observation for CPUs and GPUs is that devices of later version have higher computing ability. For example, the E5-2695 chip is on average 2.43 times faster than the E5-2650 one since it has higher clock rate and larger cache memory size. The V-100 device dated 2017 is on average 6.25 times more powerful than its K-40 counterpart because it has a total of 5120 cores available with higher clock rate and wider memory bandwidth. It could also be inferred that the development of GPU computing ability is much faster than that of CPU during recent years.

By a cross comparison, applying UGKS on GPUs could obtain better performance than on CPUs produced around same year. For example, the K-40 device is faster than the E5-2650 CPU with a maximum speedup of 18.85x on Mesh 2. The V-100 device could attain an overwhelming speedup of 118.38x for Mesh 3 and 4 while the E5-2695 chip introduced one year earlier could only acquire a maximum of 28.77x.

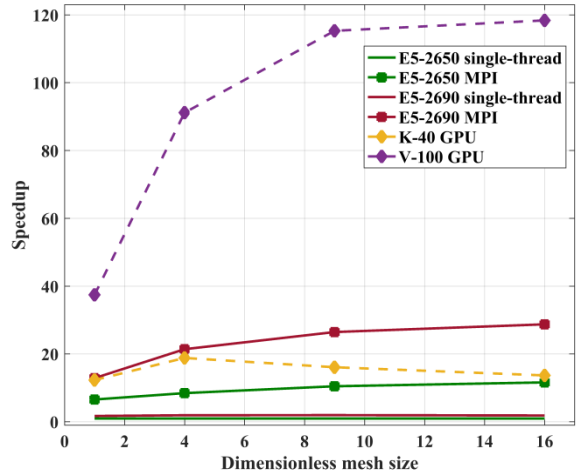

Figure 5 Performance comparisons of UGKS on CPUs and GPUs

The performances of CPUs follow the Amdahl's law as is described previously. However, the speedup plots for K-40 GPU saturates at a certain problem size and then decrease as problem size increases. (If the problem size further increases, V-100 device sees the same trend). This might be due to insufficient core number on GPU so that although a kernel function calls a full parallelization of block and thread, not all appointed blocks start at the same time. Some need to wait the completion of others before being put into calculation.

Another concern is that the speedup for K-40 device is relatively limited with maximum value of 18.85x. In terms of absolute value, it is relatively small as compared to other GPU applications with finite difference method or lattice Boltzmann models [10,12]. This issue as analyzed in in section 2.B is mainly due to the fact that the flux evaluation process in UGKS is not suitable for a full parallelization. For the sequential part and moment calculation, a compromised strategy is adopted while for the other processes, full parallelization is used.

*Influence of velocity space discretization*

29×29 grid points in velocity space might be an acceptable choice for normal gas dynamics calculations. However, as explained in the introduction section, more grid points are needed if the UGKS were to be used to model micro-sized particle dynamics. To provide performance guidance on this issue, tests with increased velocity space grid points are conducted. Fig. 6 compares influence of the velocity space discretization on speedups with E5-2650 and K-40 device for 4 kinds of mesh sizes.

For parallel algorithm with CPU, increasing velocity space grid points further boosts the accelerations. For example, with a fixed mesh size at Mesh 4, the speedup increases from 11.64x (29×29) to 20.99x (49×49) and to 21.59x (69×69). However, the K-40 GPU device provides an inversed trend compared to CPU. The speedups generally fall down when more points are added in velocity space. For instance, on Mesh 3, K-40 GPU has a speedup of 16.11x with 29×29 grid points, but it reduces to 15.60x and 13.44x with 49×49 and 69×69 grid points respectively. The descending performance of GPU device might be related to the limited computing resources within one GPU card.

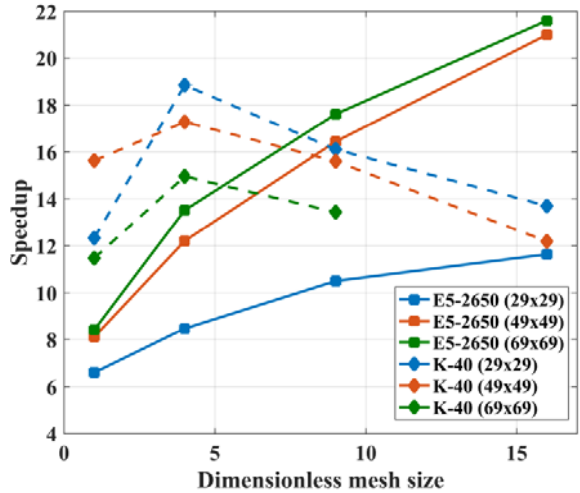

Figure 6 Influence of particle velocity space discretization
\* K-40 GPU runs out of memory for Mesh 4 with 69×69 velocity space grid points

## 4. Conclusion

Primitive parallel algorithms on CPU and GPU are implemented for the Unified Gas-Kinetic Scheme. Their performances are tested and compared in detail by a two dimensional channel flow case with 4 different mesh sizes and 3 different molecular velocity space discretizations. The parallel MPI-CPU algorithm adopts a simple one dimensional block partition on spatial space. A compromised two level parallelization strategy is applied to the algorithm for GPU based on the intrinsic feature of UGKS formulation.

The results show that the parallel MPI algorithm for CPU could accelerate the two dimensional UGKS calculation following the Amdahl's law. Depending on size of the problem, the maximum speedups could reach 11.64x for E5-2650 device and 15.23x for E5-2695 device. GPU, on the other side, is also able to significantly boost parallel computation. For the tested case, the K-40 GPU could obtain a maximum speedup of 18.85x while the latest V-100 GPU could reach a 118.38x speedup.

For UGKS with large grid point number in molecular velocity space, comparison has been conducted between E5-2560 CPU and K-40 GPU. The MPI algorithm on CPU has a predictable increased performance with increased velocity grid points. However, the algorithm on GPU device shows a descending trend with it due to a lack of computing resources within only one GPU card.

## Acknowledgement

The authors would like to thank the Research Computing Service of Old Dominion University for providing GPU resources for the current study.